\documentclass[twocolumn,prl,superscriptaddress,showpacs,10pt,aps,draft]{revtex4-1} 

\usepackage{color}

\usepackage{amsmath}
\usepackage{amssymb}
\usepackage{graphicx}
\usepackage{color}

\usepackage[T1]{fontenc}
\usepackage[utf8]{inputenc}

\usepackage{amsthm}
\usepackage{bbm}
\begin{document}
\title{Comment on ``Noise and Disturbance in Quantum Measurements: An Information-Theoretic Approach''} 

\author{Paul Busch}
\email{paul.busch@york.ac.uk}
\affiliation{Department of Mathematics, University of York, York, United Kingdom}

\author{Pekka Lahti}
\email{pekka.lahti@utu.fi}
\affiliation{Turku Centre for Quantum Physics, Department of Physics and Astronomy, University of Turku, FI-20014 Turku, Finland}

\author{Reinhard F. Werner}
\email{reinhard.werner@itp.uni-hannover.de}
\affiliation{Institut f\"ur Theoretische Physik, Leibniz Universit\"at, Hannover, Germany}

\date{\today}
%\begin{abstract}

%\end{abstract}
\maketitle
In their  interesting paper \cite{Buscemi2014}, Buscemi {\em et al} derive a state-independent entropic trade-off relation for the noise (approximation error) in the measurement of one observable and the necessary disturbance imparted thereby on another observable. We appreciate that this work is very much in the spirit of our recent letter \cite{BLW2013c}, where we have derived two variants of error-disturbance relations for position and momentum in terms of calibration errors and worst case quadratic deviations, respectively. Yet, the paper \cite{Buscemi2014} contains some comments on our letter that are either misleading or incorrect. Crucially, there is an internal tension in the presentation that betrays a misrepresentation of the research programme underlying both papers \cite{Buscemi2014} and \cite{BLW2013c} which we feel needs correcting lest it be perpetuated by repetition.

In the conclusion of their work, the authors of \cite{Buscemi2014} describe their results concerning state-independent entropic measurement uncertainty relations as fundamental, and we agree with this assessment. However, the discussion implies that they consider the significance of their relation to be restricted to the case of discrete observables. They do consider an extension to continuous observables, such as position and momentum, but immediately declare this to be ``purely formal, with no operational counterparts''. This verdict is then also applied to our calibration error relation. Based on this ``formality'' claim, for which no justification is offered other than the hint that continuous observables do not have proper eigenstates, the authors conclude that ``it appears that state-dependent noise-disturbance and joint-measurement relations \dots\ may be preferable for continuous observables.'' 

In contrast, we maintain that both approaches to formulating measurement uncertainty relations -- with state independent or state dependent error measures -- have their separate uses and merits, and that there is
no reason why either should be limited to a certain type of observables %We have explained the reasons in 
%great detail in our recent critical analysis of different proposed error measures in 
\cite{BLW2013a}. 
State dependent error and disturbance measures are useful if, for example, one wishes to carry out information theoretic tasks in which the disturbance should be limited in the case of a specific state. An error-disturbance relation would then tell us how the accuracy in a measurement must be limited. By contrast, state-independent error measures, which can be defined as suitable mean or worst-case errors, are suitable as figures of merit for a measuring device. The corresponding uncertainty relations describe the limitations that
all possible devices are subjected to if they are to be used for jointly approximating a pair of incompatible quantities. Unfortunately, the recent hype about alleged violations of Heisenberg's error-disturbance relation has distracted somewhat from appreciating the important role played by state-independent measures in quantifying this fundamental measurement limitation. 

Moreover, it is incorrect to say that calibration error measures are without operational counterpart or content.
In fact, it is standard experimental practice to calibrate a measuring device by applying it to situations where
the property to be measured has a fairly sharply determined, known value or distribution, and then comparing this value or distribution with the device's output distribution. This is exactly what is being captured with the
error measures defined in both \cite{BLW2013c} and \cite{Buscemi2014}. These measures may be difficult to implement practically; but that does not make them void of operational meaning.

Incidentally, the claim of \cite{Buscemi2014} that their relation $V^Q_{\mathsf N}V^P_{\mathsf D}\ge\hbar^2/4$ is stronger than ours is logically unfounded: their $V$ quantities are defined in terms of very specific families of approximate position  and momentum eigenstates while our measures take into account all possible sufficiently localized states. There is therefore no direct comparison possible between our respective quantities.
Nevertheless it is gratifying to see that \cite{Buscemi2014} attempts to strengthen our error-disturbance relation, considering that in a recent arXiv publication one of its coauthors attempted to disprove it \cite{Ozawa2013,BLW2014a}.

%\bibliographystyle{unsrt}
%\bibliography{UR}

\end{document}